\documentclass{article}
\usepackage{amsmath,amsthm}
\usepackage{amssymb,latexsym}
\usepackage[mathscr]{eucal}
\usepackage{setspace}
\usepackage{graphics}
\usepackage{array}

\newcommand{\pa}{\partial}

\newcommand{\na}{\nabla}
\newcommand{\ti}{\times}
\newcommand{\bs}{\boldsymbol}
\newcommand{\lp}{\left(}
\newcommand{\rp}{\right)}
\newcommand{\lb}{\left[}
\newcommand{\rb}{\right]}
\newcommand{\lc}{\left\{}
\newcommand{\rc}{\right\}}
\newcommand{\la}{\left<}
\newcommand{\ra}{\right>}
\newcommand{\be}{\begin{equation}}
\newcommand{\ee}{\end{equation}}

\newcommand{\ihat}{\bf\hat{i}}

\begin{document}

\title{\textbf{The Okubo-Weiss Criteria in Two-Dimensional Hydrodynamic and Magnetohydrodynamic Flows}}         
\author{B. K. Shivamoggi\footnote{Permanent Address: University of Central Florida, Orlando, FL 32816-1364, USA}, G. J. F. van Heijst and L. P. J. Kamp ~\\
J. M. Burgers Centre and Fluid Dynamics Laboratory\\
Department of Physics\\
Eindhoven University of Technology\\
NL-5600MB Eindhoven, The Netherlands
}        
\date{}          
\maketitle

\large{\bf Abstract}

The Okubo \cite{Oku}-Weiss \cite{Wei} criterion is recast by using the 2D hydrodynamic Beltrami condition (Shivamoggi et al. \cite {Shi}) that approximates the {\it slow flow-variation ansatz} imposed in its derivation. This turns out to provide an interesting interpretation of the Okubo-Weiss criterion very logically in terms of the topological properties of the underlying vorticity manifold. These developments are then extended to 2D quasi-geostrophic flows (via the {\it potential divorticity} framework) and magnetohydrodynamic flows and the Okubo-Weiss criteria for these cases are considered.

\pagebreak

\noindent\Large\textbf{1. Introduction}\\

\large  A central question in the problem of transport in two-dimensional (2D) turbulent flows is how to divide a vorticity field into hyperbolic (cascading turbulence) and elliptic (coherent vortex) regions because the topology of 2D turbulence is parameterized in terms of the relative dominance of flow deformation or flow rotation.\footnote{It is of interest to note that, in considering the shape of a material curve in periodic 2D incompressible flows, Berry et al. \cite{Ber} distinguished the elliptic and hyperbolic regions via the wrapping-around action (termed a ``{\it whorl}") in the former and a stretching-compressing action (termed a ``{\it tendril}") in the latter.} Okubo \cite{Oku} and Weiss \cite{Wei} gave a kinematic criterion to serve as a diagnostic tool towards this goal which has been widely used in numerical simulations (Brachet et al. \cite{Bra}, Ohkitani \cite{Ohk}, Babiano and Provenzale \cite{Bab}) and laboratory experiments (Ouelette and Gollub \cite{Oue}) of 2D hydrodynamic flows.\footnote{The Okubo-Weiss parameter describing the local strain-vorticity balance in the horizontal flow field of a shallow fluid layer turns out also to quantify the deviations from two-dimensionality of this flow (Balkovsky et al. \cite{Bal}, Cieslik et al. \cite{Cie}). More specifically, the Okubo-Weiss parameter turns out to be the source term in the Poisson equation for the pressure Hessian matrix (Kamp \cite{Kam}).} A key assumption underlying the Okubo-Weiss criterion is that the vorticity gradient field evolves adiabatically with respect to the underlying straining flow-velocity gradient field which is assumed to evolve temporally  slowly. This issue was explored by Basdevant and Philipovitch \cite{Bas} who tried to improve on it by invoking the topological properties of the pressure field, while Hua and Klein \cite{Hua} tried to include the strain-rate time evolution explicitly. In this paper, the Okubo-Weiss criterion is recast by using the 2D hydrodynamic Beltrami condition (Shivamoggi et al.  \cite{Shi}) that approximates the {\it slow flow-variation ansatz} imposed in its derivation. This turns out to provide an interesting interpretation of the Okubo-Weiss criterion very logically in terms of the topological properties of the underlying vorticity manifold. These developments are then extended to 2D quasi-geostrophic flows (via the {\it potential divorticity} framework) and magnetohydrodynamic (MHD) flows and the Okubo-Weiss criteria for these cases are considered.

\vspace{.3in}

\noindent\Large\textbf{2. Recasting the Okubo-Weiss Criterion via the Beltrami Condition }\\ 

\large The vorticity dynamics in 2D hydrodynamic flows is governed by the following equation (Kida \cite{Kid}, Kuznetsov et al. \cite{Kuz}),
\be\tag{1a}
\frac{\pa \bs{\mathscr{B}}}{\pa t} = \na \ti \lp {\bf v} \ti \bs{\mathscr{B}} \rp
\ee
or
\be\tag{1b}
\frac{D \bs{\mathscr{B}}}{D t} \equiv \frac{\pa \bs{\mathscr{B}}}{\pa t} + \lp {\bf v} \cdot \na \rp \bs{\mathscr{B}} = \lp \bs{\mathscr{B}} \cdot \na \rp {\bf v}
\ee
where ${\bf v} = \la u, v \ra$ is the flow velocity, and $\bs{\mathscr{B}}$ is the divorticity,
\be\tag{2}
\bs{\mathscr{B}} \equiv \na \ti \bs{\omega}, ~\bs{\omega} \equiv \na \ti {\bf v}.
\ee

Equation (1b) may be rewritten as
\be\tag{1c}
\frac{D \bs{\mathscr{B}}}{D t} = \bs{\mathscr{A}} \cdot \bs{\mathscr{B}}
\ee
where $\bs{\mathscr{A}}$ is the velocity gradient matrix,
\be\notag
\bs{\mathscr{A}} \equiv \lb
\begin{matrix}
\pa u/\pa x & \pa u/\pa y\\
\pa v/\pa x & \pa v/\pa y
\end{matrix}
\rb = \frac{1}{2} \lb
\begin{matrix}
s_1 & s_2 - \omega\\
s_2 + \omega & -s_1
\end{matrix}
\rb
\ee
\be\tag{3}
s_1 \equiv -2 \frac{\pa v}{\pa y}, ~s_2 \equiv \frac{\pa v}{\pa x} + \frac{\pa u}{\pa y}, ~\omega \equiv \frac{\pa v}{\pa x} - \frac{\pa u}{\pa y}.
\ee

If the straining flow-velocity gradient tensor $\na {\bf v}$ is assumed, following Okubo \cite{Oku} and Weiss \cite{Wei}, to temporally evolve {\it slowly} so the divorticity field evolves adiabatically with respect to the straining flow-velocity gradient field \footnote{This {\it ansatz} is also implicit in Toda's criterion (Toda \cite{Toda}) to predict the onset of chaotic motion which is based on a linear stability of trajectories in the neighborhood of the reference trajectory---the stability properties of neighboring trajectories are assumed to follow the reference trajectory adiabatically.}, equation (1c) may be locally approximated by an eigenvalue problem with eigenvalues given by
\be\tag{4}
\lambda^2 =u_y v_x+v_y^2= \frac{1}{4} \lp {s_1}^2 + {s_2}^2 - \omega^2 \rp \equiv Q.
\ee
The Okubo-Weiss parameter $Q$ is a measure of the relative importance of flow strain \\($Q$ $>$ 0, hyperbolic) and vorticity ($Q$ $<$ 0, elliptic). Numerical simulations (Brachet et al. \cite{Bra}, Ohkitani \cite{Ohk}, Babiano and Provenzale \cite{Bab}) and laboratory experiments (Ouellette and Gollub \cite{Oue}) of 2D hydrodynamic flows confirmed that coherent vortices are indeed located in elliptic regions while divorticity sheets are located\footnote{It may be noted that divorticity sheets are also more likely to occur near vorticity nulls due to selective rapid viscous decay of vorticity in these layers (Shivamoggi et al. \cite{Shi}), just as vortex sheets are more likely to form near velocity nulls in 3D hydrodynamic flows.} in hyperbolic regions.

An interesting interpretation of the Okubo-Weiss parameter very logically in terms of the topological characteristics of the underlying vorticity manifold becomes available by noting that the {\it slow flow-variation ansatz} used above may be approximated by the Beltrami condition\footnote{A similar approach was taken previously (Shivamoggi and van Heijst \cite{Shi3}) in dealing with the ``{\it slow variation}" restriction used in Flierl-Stern-Whitehead \cite{Fli} {\it zero angular momentum theorem} for localized nonlinear structures in 2D quasi-geostrophic flows on the $\beta$-plane.} for 2D hydrodynamics (Shivamoggi et al. \cite{Shi}) governed by equations (1a-c),

\be\tag{5}
\bs{\mathscr{B}} = a {\bf v}
\ee
$a$ being an arbitrary constant. Using (5), the Okubo-Weiss parameter $Q$ may be recast as follows, 

\be\tag{6}
Q = \frac{1}{a^2} \lb \lp \frac{\pa^2 \omega}{\pa x \pa y} \rp^2 - \frac{\pa^2 \omega}{\pa x^2} \frac{\pa^2 \omega}{\pa y^2} \rb.
\ee

(6) implies that the Okubo-Weiss parameter also characterizes the topological properties of the vorticity manifold - it is in fact, to within a positive multiplicative factor (see Footnote 6) the negative of the {\it Gaussian} curvature of the vorticity surface.\footnote{The Gaussian curvature of a vorticity surface $\omega=\omega\lp x,y\rp$ is
$$\kappa=\frac{\omega_{xx}\omega_{yy}-\omega^2_{xy}}{\lp 1+\omega^2_x+\omega^2_y\rp^2}.$$} Thus, the character of the ensuing  2D flow behavior appears to be rooted in the local topological properties of the underlying  vorticity manifold. It may be mentioned that the reduction (6) was pointed out by Larcheveque \cite{Lar} on the premise of replacing streamlines by isovorticity lines which lacked, as Larcheveque \cite{Lar} admitted, any dynamical meaning - streamlines are actually isomorphic to divorticity lines (as implied by the Beltrami condition (5)) rather than the isovorticity lines.

\vspace{0.15in}

\noindent\Large\textbf{3. Reformulation in Polar Coordinates}\\ 

\large In plane polar coordinates $\lp r,\theta\rp$, equation (1b) written in the component form  is 

\be\tag{7a}
\displaystyle\frac{\pa B_r}{\pa t}+v_r\frac{\pa B_r}{\pa r}+v_\theta\frac{\pa B_r}{r\pa\theta}-\frac{v_\theta B_\theta}{r}=B_r\frac{\pa v_r}{\pa r}+B_\theta\frac{\pa v_r}{r\pa \theta}-\frac{B_\theta v_\theta}{r}
\ee

\vspace{0.05in}

\be\tag{7b}
\displaystyle\frac{\pa B_\theta}{\pa t}+v_r\frac{\pa B_\theta}{\pa r}+v_\theta\frac{\pa B_\theta}{r\pa\theta}+\frac{v_\theta B_r}{r}=B_r\frac{\pa v_\theta}{\pa r}+B_\theta\frac{\pa v_\theta}{r\pa \theta}+\frac{B_\theta v_r}{r}
\ee

\noindent so the velocity gradient matrix $\bs{\mathscr{A}}$ in equation (1c) becomes

\be\tag{8a}
\bs{\mathscr{A}} \equiv \lb
\begin{matrix}
\begin{aligned}
\displaystyle\frac{\pa v_r}{\pa r}\phantom{xx}&\displaystyle\frac{\pa v_r}{r\pa\theta}&-\phantom{x} \displaystyle\frac{v_\theta}{r}\\
\\
\displaystyle\frac{\pa v_\theta}{\pa r}\phantom{xx}&\displaystyle\frac{\pa v_\theta}{r\pa\theta}&+\phantom{x}\displaystyle\frac{v_r}{r}
\end{aligned}
\end{matrix}
\rb = \frac{1}{2} \lb
\begin{matrix}
s_1 & s_2 - \omega\\
s_2 + \omega & -s_1
\end{matrix}
\rb
\ee

\noindent where,

\be\tag{8b}
\displaystyle s_1\equiv 2\frac{\pa v_r}{\pa r},\phantom{xx} s_2\equiv r \frac{\pa}{\pa\theta}\lp\frac{v_\theta}{r}\rp+\frac{\pa v_r}{r\pa\theta},~\omega\equiv\frac{1}{r}\frac{\pa}{\pa r}\lp r v_\theta\rp-\frac{\pa v_r}{r\pa\theta}.\ee

On assuming again that the divorticity field evolves adiabatically with respect to the straining flow-velocity gradient field, equation (1c) may be locally approximated by an eigenvalue problem with eigenvalues given by

\be\tag{9a}
\lambda^2=\frac{1}{4}\lp s_1^2+s_2^2-\omega^2\rp=\lp\displaystyle\frac{\pa v_r}{r\pa\theta}-\frac{v_\theta}{r}\rp \lp\frac{\pa v_\theta}{\pa r}\rp-\lp\frac{\pa v_r}{\pa r}\rp\lp\frac{\pa v_\theta}{r\pa\theta}+\frac{v_r}{r}\rp\equiv Q.
\ee  

\noindent On using the mass-conservation equation, 

\be\notag
\displaystyle\frac{1}{r}\displaystyle\frac{\pa}{\pa r}\lp r v_r\rp+\displaystyle\frac{\pa v_\theta}{r\pa\theta}=0
\ee

\noindent (9a) may be written alternatively as

\be\tag{9b}
\displaystyle \lambda^2=\lp\frac{\pa v_r}{\pa r}\rp^2+ \displaystyle \frac{1}{r}\frac{\pa v_r}{\pa\theta}\frac{\pa v_\theta}{\pa r}-\frac{1}{r}v_\theta\frac{\pa v_\theta}{\pa r}\equiv Q\ee

\noindent or

\be\tag{9c}
\displaystyle \lambda^2=\lp\frac{\pa v_\theta}{\pa r}\rp \displaystyle\lp \frac{\pa v_r}{r\pa \theta}-\frac{ v_\theta}{r}\rp+\lp\frac{v_r}{r}+\frac{\pa v_\theta}{r\pa\theta}\rp^2\equiv Q.
\ee

\noindent On using the Beltrami condition (5) for 2D hydrodynamics, (9c) becomes

\be\tag{10}
Q=\displaystyle\frac{1}{a^2r^4}\lb-r^2\frac{\pa^2\omega}{\pa r^2}\lp\frac{\pa^2\omega}{\pa\theta^2}+r \frac{\pa \omega}{\pa r}\rp+\lp\frac{\pa\omega}{\pa\theta}-r\frac{\pa^2\omega}{\pa r\pa \theta}\rp^2\rb
\ee

\noindent which to within a positive multiplicative factor is the negative of the {\it Gaussian} curvature of the vorticity surface.\footnote{The Gaussian curvature of a vorticity surface $\displaystyle\omega=\omega\lp r,\theta\rp$ is

$$ \displaystyle\kappa=\displaystyle\frac{r^2\displaystyle\frac{\pa^2\omega}{\pa r^2}\lp\displaystyle\frac{\pa^2\omega}{\pa\theta^2}+r\displaystyle\frac{\pa\omega}{\pa r}\rp-\lp\displaystyle\frac{\pa\omega}{\pa\theta}-r\displaystyle\frac{\pa^2\omega}{\pa r\pa\theta}\rp^2}{\lb r^2\lc\lp\displaystyle\frac{\pa\omega}{\pa r}\rp^2+1\rc+\lp\displaystyle\frac{\pa\omega}{\pa\theta}\rp^2\rb^2}$$}

\vspace{0.10in}
\large{\bf Example 1:} As an example of the above formulation, consider an axisymmetric flow with stream function given by

\be\tag{11}
\psi=\psi\lp r\rp
\ee

\noindent so the flow velocity is given by

\be\tag{12}
v_r=0,~v_\theta=-\frac{d\psi}{dr}.
\ee

Using (12), (8b) becomes

\be\tag{13}
\displaystyle s_1=0,~ s_2=-r \frac{d}{dr}\lp\frac{1}{r}\frac{d\psi}{dr}\rp,~\omega=-\frac{1}{r}\frac{d}{dr}\lp r\frac{d\psi}{dr}\rp.
\ee

Using (13), (9b) gives

\be\tag{14}
\lambda^2=-\frac{1}{r}\frac{d\psi}{dr}\frac{d^2\psi}{dr^2}
\ee

\noindent which agrees with the result given by Lapeyre et al. \cite{Lap}.

For a flow resembling a rigid-body rotation, 

\be\tag{15}
\psi=-\frac{1}{2}\Omega r^2.
\ee

\noindent (14) becomes

\be\tag{16}
\lambda^2=-\Omega^2<0
\ee

\noindent as to be expected.
 
\vspace{0.10in}
\large{\bf Example 2:} Consider the Lamb-Oseen vortex, 

\be\tag{17}
v_\theta=\frac{\Gamma}{2\pi r}\lp 1-e^{-r^2/4\nu t}\rp.
\ee

\noindent $\Gamma$ being the total circulation associated with the vortex.

Using (17), (9b) gives

\be\tag{18}
\lambda^2=-\frac{\Gamma^2}{4\pi^2 r^3}\lp 1-e^{-r^2/4\nu t}\rp\lb-\frac{1}{r^2}\lp 1-e^{-r^2/4\nu t}\rp+\frac{1}{2\nu t}e^{-r^2/4\nu t}\rb
\ee

\noindent which leads to 

\be\tag{19}
\lambda^2\approx
\lc
\begin{matrix}
\begin{aligned}
-\displaystyle\frac{\Gamma^2}{64\pi\nu^2t^2},~~&r<<\sqrt{4\nu t}\\
\\
\displaystyle\frac{\Gamma^2}{4\pi r^4},~~&r>>\sqrt{4\nu t}
\end{aligned}
\end{matrix}\right.\ee

\noindent signifying an elliptic region inside the vortex core and a hyperbolic region away from the vortex core.

\vspace{0.10in}
\large{\bf Example 3:} Consider the Burgers vortex,\footnote{Burgers vortex describes the interplay between the intensification of vorticity due to the imposed straining flow and the diffusion of vorticity due to the action of viscosity.}

\be\tag{20}
v_r=-\frac{a}{2}r, ~ v_\theta=\frac{\Gamma}{2\pi r}\lp 1-e^{-ar^2/4\nu}\rp\approx\lc\begin{matrix}\begin{aligned}\lp\frac{\Gamma a}{8\pi\nu}\rp r,~& r<<\sqrt{\frac{4\nu}{a}}\\\frac{\Gamma}{2\pi r},~& r>>\sqrt{\frac{4\nu}{a}}\end{aligned}\end{matrix}\right.,~v_z=az\ee  

Using (20), (8a) gives

\be\tag{21}
\triangle\equiv|\bs{\mathscr{A}}|=
\lc
\begin{matrix}
\begin{aligned}
\displaystyle\frac{a^2}{4}+\frac{\Gamma^2 a^2}{64\pi^2\nu^2},~~&r<<\sqrt{\frac{4\nu}{a}}\\
\\
\displaystyle\frac{a^2}{4},~~&r>>\sqrt{\frac{4\nu}{a}}
\end{aligned}
\end{matrix}\right.\ee

The eigenvalues of the volocity gradient matrix are given by

\be\tag{22}
\lambda^2+a\lambda+\triangle=0
\ee

\noindent from which, on assuming $\Gamma >>4\pi\nu$, we obtain

\be\tag{23}
\lambda^2=
\lc
\begin{matrix}
\begin{aligned}
\displaystyle-\frac{\Gamma^2 a^2}{64\pi^2\nu^2},~~&r<<\sqrt{\frac{4\nu}{a}}\\
\\
\displaystyle\frac{a^2}{4},~~&r>>\sqrt{\frac{4\nu}{a}}
\end{aligned}
\end{matrix}\right.\ee

\noindent signifying again an elliptic region inside the vortex core and a hyperbolic region away from the vortex core.

\vspace{0.10in}
\newpage
\noindent\Large\textbf{4. The Okubo-Weiss Criterion for Quasi-geostrophic Flows}\\

\large Consider a 2D quasi-geostrophic\footnote{Quasi-geostrophic dynamics refers to the nonlinear dynamics governed by the first-order departure from the linear geostrophic balance between the Coriolis force and pressure gradient transverse to the rotation axis of a rapidly rotating fluid (Charney \cite{Cha}).} flow in which the baroclinic effects are produced by the deformed free surface of the ocean. The governing equation (in appropriate units) is (Charney \cite{Cha})

\be\tag{24}
\frac{\pa {\bf q}}{\pa t} + \lp {\bf v} \cdot \na \rp {\bf q} = 0
\ee
where {\bf q} is the potential vorticity vector,
\be\tag{25}
{\bf q} \equiv \bs{\omega} - k^2 \bs{\psi} + {\bf f}
\ee
{\bf f} is the Coriolis parameter, ${\bf f} = \la 0, 0, f(y) \ra$ ($x$ is along the local east and $y$ is along the local north direction), $k$ is the inverse Rossby radius of deformation, $k \equiv \sqrt{{f_0}^2/gH}$, ($f_0$ being the local value of $|{\bf f}|$ and H the depth of the ocean, which is taken to be uniform), and
\be\tag{26}
{\bf v} \equiv -\na \ti \bs{\psi}
\ee
as per the fluid incompressibility condition. We are using the simplest mathematical model of large-scale, nearly horizontal oceanic motion incorporating the force of gravity and the Coriolis force due to Earth's rotation, which is the one-layer homogeneous ocean with a uniform depth and spherical free surface.

Upon taking the curl of equation (24), we obtain
\be\tag{27a}
\frac{\pa \bs{\mathscr{D}}}{\pa t} = \na \ti \lp {\bf v} \ti \bs{\mathscr{D}} \rp
\ee
where $\bs{\mathscr{D}}$ is the {\it potential divorticity} vector (in analogy to the potential vorticity vector {\bf q}),
\be\tag{28}
\bs{\mathscr{D}} \equiv \na \ti {\bf q} = \bs{\mathscr{B}} + k^2 {\bf v} + {\bf h}
\ee
and in the $\beta$-plane approximation\footnote{The $\beta$-plane approximation corresponds to replacing the curved surface of the earth locally by a tangent plane, but allowing the Coriolis parameter {\bf f} to vary linearly with latitude (the $y$-direction).},
\be\tag{29}
{\bf f} = \la 0, 0, f_0 + \beta y \ra
\ee
we have
\be\tag{30}
{\bf h} \equiv \na \ti {\bf f} = \la \beta, 0, 0 \ra
\ee
$\beta$ being the planetary vorticity gradient.

Equation (27a) may be rewritten as
\be\tag{27b}
\frac{D \bs{\mathscr{D}}}{D t} = \bs{\mathscr{A}} \cdot \bs{\mathscr{D}}.
\ee

Following Okubo \cite{Oku} and Weiss \cite{Wei}, and assuming that the potential divorticity field evolves adiabatically with respect to the straining flow-velocity gradient field, equation (10b) may again be locally approximated by an eigenvalue problem with eigenvalues given by,
\be\tag{31}
\lambda^2 = u_y v_x + {v_y}^2 \equiv Q.
\ee

Equation (27a) yields for the Beltrami state,
\be\tag{32}
\bs{\mathscr{D}} = b {\bf v}
\ee
$b$ being an arbitrary constant. Using (32), the Okubo-Weiss parameter $Q$ becomes
\be\tag{33}
Q \equiv \frac{1}{b^2} \lb \lp \frac{\pa^2 \omega}{\pa x \pa y} \rp^2 - \frac{\pa^2 \omega}{\pa x^2} \frac{\pa^2 \omega}{\pa y^2} \rb
\ee
which is the same as (6) for 2D hydrodynamic case. This shows that the Okubo-Weiss parameter $Q$ is robust and remains intact under extension to 2D quasi-geostrophic flows (in the $\beta$-plane approximation to the Coriolis parameter). The inclusion of the nonlinear variation in the Coriolis parameter (the so-called $\gamma${\it -effect}, which becomes important in the polar region)\footnote{Upon including the $\gamma$-effect, (29) becomes
\be\notag
{\bf f} = \la 0, 0, f_0 + \beta y + \frac{\gamma}{2} y^2 \ra .
\ee} will, however, lead to changes in the Okubo-Weiss parameter,
\be\tag{34}
Q = \frac{1}{b^2} \lb \lp \frac{\pa^2 \omega}{\pa x \pa y} \rp^2 - \frac{\pa^2 \omega}{\pa x^2} \lp \frac{\pa^2 \omega}{\pa y^2} + \gamma \rp \rb.
\ee

\vspace{.3in}

\noindent\Large\textbf{5. The Okubo-Weiss Criterion for MHD Flows}\\

\large In the MHD model, the dynamics is dominated by ions with electrons serving to shield out rapidly any charge imbalances. Consider a 2D incompressible MHD flow. The equation governing the transport of the magnetic field ${\bf B} = \la B_1, B_2 \ra$\footnote{Equation (35a) follows from Ohm's law for an infinitely-conducting fluid,
\be\notag
{\bf E} + \frac{1}{c} {\bf v} \times {\bf B} = {\bf 0}
\ee
and Faraday's law,
\be\notag
- \frac{1}{c} \frac{\partial {\bf B}}{\partial t} = \nabla \times {\bf E}
\ee
$c$ being the velocity of light and {\bf E} being the electric field.} (or Ohm's law) is (Goedbloed and Poedts \cite{Goe}), in usual notation, 
\be\tag{35a}
\frac{\partial {\bf B}}{\partial t} = \nabla \times ( {\bf v} \times {\bf B} )
\ee
or
\be\tag{35b}
\frac{D {\bf B}}{D t} = \lp {\bf B} \cdot \na \rp {\bf v}
\ee
which may be rewritten as
\be\tag{35c}
\frac{D {\bf B}}{D t} = \bs{\mathscr{A}} \cdot {\bf B}.
\ee

If we now assume that the magnetic field evolves adiabatically with respect to the straining flow velocity gradient field, equation (35c) may again be locally approximated by an eigenvalue problem with eigenvalues given by,
\be\tag{36}
\lambda^2 = u_y v_x + {v_y}^2 \equiv Q.
\ee

Noting that the MHD Beltrami state (Shivamoggi \cite{Shi2}), from equation (35a), corresponds to the so-called Alfv\'{e}nic state (Hasegawa \cite{Has}),
\be\tag{37}
{\bf B} = a {\bf v}
\ee
$a$ being an arbitrary constant, the Okubo-Weiss parameter $Q$ for the MHD case becomes
\be\tag{38}
Q = \frac{1}{a^2} \lp\displaystyle B_{1y} {B_2}_x + {B^2_{2y}} \rp=\frac{1}{4a^2}\lp b_1^2+ b_2^2- J^2/c^2\rp.
\ee
\noindent where,

$$b_1\equiv-2B_{2y}, ~b_2\equiv{B_2}_x+B_{1y},~\frac{J}{c}\equiv B_{2x}-B_{1y}.$$

So, the Okubo-Weiss parameter $Q$ is a measure of the relative importance of magnetic shear $(Q>0$, hyperbolic) and electric current $(Q<0,$ ~elliptic). This result, in conjunction with (36) and (37), is in accord with the eigenvalues of the $\nabla{\bf B}$-matrix becoming purely imaginary near the $O$-points or real near the $X$-points of the magnetic field lines (Greene \cite{Gre}).

In terms of the magnetic vector potential {\bf A} given by
\be\tag{39}
{\bf B} \equiv \na \ti {\bf A}, ~{\bf A} = A \ {\ihat}_z
\ee
(38) becomes
\be\tag{40}
Q = \frac{1}{a^2} \lb \lp \frac{\pa^2 A}{\pa x \pa y}\rp^2 - \frac{\pa^2 A}{\pa x^2} \frac{\pa^2 A}{\pa y^2} \rb.
\ee
(40) implies that the Okubo-Weiss parameter $Q$ for the MHD case characterizes the topological properties of the magnetic flux surface - it is the negative of the {\it Gaussian} curvature of the magnetic flux surface.  As with the case of 2D hydrodynamic flows, (23) can therefore serve as a useful diagnostic tool to parameterize the magnetic field topology in 2D MHD flows.

\large{\bf Example 4:} As an example of the above formulation, consider a MHD flow with the magnetic flux function given by

\be\tag{41}
A\lp x,y\rp=\frac{k}{2}\lp \alpha x^2-y^2\rp.
\ee

\noindent So, the magnetic field is given by the hyperbolic/elliptic configuration near an $X\lp O\rp$-type $\lp\alpha\gtrless 0\rp$ magnetic neutral point, 

\be\tag{42}
B_1=-ky,~B_2=-k\alpha x.
\ee

\noindent Thus, 

\be\tag{43}
b_1=0,~ b_2=-k\lp 1+\alpha\rp,~\frac{J}{c}=k\lp 1-\alpha\rp
\ee

\noindent which shows that the current density $J=0$, unless $\alpha\not= 1$. However, the magnetic field topology is determined only by whether $\alpha\lessgtr 0$, as shown below.

Using (43), (38) gives

\be\tag{44}
Q=\lp \frac{k^2}{a^2}\rp \alpha \gtrless 0, ~\alpha \gtrless 0
\ee

\noindent as to be expected.
\vspace{0.3in}

In plane-polar coordinates, a reformulation similar to that in Section 3, leads, in place of (38), to 

\be\tag{45a}
Q=\frac{1}{a^2}\lb\lp\frac{\pa B_r}{\pa r}\rp^2+\frac{1}{r}\frac{\pa B_r}{\pa\theta}\frac{\pa B_\theta}{\pa r}-\frac{1}{r}B_\theta\frac{\pa B_\theta}{\pa r}\rb.
\ee

\noindent On using the Gauss law, 

\be\tag{46}
\frac{1}{r}\frac{\pa}{\pa r}\lp r B_r\rp +\frac{\pa B_\theta}{r\pa\theta}=0
\ee

\noindent (45a) may be alternatively expressed as 
\be\tag{45b}
Q=\frac{1}{a^2}\lb\lp\frac{\pa B_\theta}{\pa r}\rp\lp\frac{\pa B_r}{r\pa\theta}-\frac{B_\theta}{r}\rp+\lp\frac{B_r}{r}+\frac{\pa B_\theta}{r\pa\theta}\rp^2\rb.
\ee
\vspace{.05in}
\large{\bf Example 5:} As an example, consider an axisymmetric MHD flow with magnetic flux function given by

\be\tag{47}
A=A\lp r\rp
\ee

\noindent so the magnetic field is given by

\be\tag{48}
B_r=0,~ B_\theta=-\frac{dA}{dr}.
\ee

Using (48), (45a) gives

\be\tag{49}
Q=-\frac{1}{r}\frac{dA}{dr}\frac{d^2A}{dr^2}
\ee

For a magnetic field generated by a uniform, unidirectional current ${\bf J}=J{\bf\hat{i}}_z$, 

\be\tag{50}
A=-\frac{1}{4c}Jr^2
\ee

\noindent (49) becomes, 

\be\tag{51}
Q=-\frac{J^2}{4c^2}<0
\ee

\noindent as to be expected. 

Consider next a magnetic field generated by a current-carrying cylinder, 

\be\tag{52}
J=J_0h\lp r_0-r\rp
\ee

\noindent$(h\lp x\rp$ being the unit step function), we have, 

\be\tag{53}
A=
\lc
\begin{matrix}
\begin{aligned}
\displaystyle-\frac{1}{2}\frac{J_0r^2}{c},~~&r<r_0\\
\\
\displaystyle-\frac{J_0r_0^2}{c}ln \lp r/r_0\rp,~~&r>r_0
\end{aligned}
\end{matrix}\right.\ee

\noindent so the magnetic field is given by

\be\tag{54}
B_\theta=
\lc
\begin{matrix}
\begin{aligned}
\displaystyle\frac{J_0r}{c},~~&r<r_0\\
\\
\displaystyle\frac{J_0r_0^2}{rc}, ~~&r>r_0
\end{aligned}
\end{matrix}\right.\ee

\noindent which represents an MHD Rankine-type flux tube.

Using (54), (45a) gives
\be\tag{55}
Q=
\lc
\begin{matrix}
\displaystyle-\frac{J_0^2}{a^2c^2},~r<r_0\\
\\
\displaystyle\frac{J_0^2r_0^4}{c^2a^2r^4},~r>r_0
\end{matrix}\right.\ee

\noindent signifying elliptic region inside the current-carrying cylinder and hyperbolic region outside it, as to be expected. 

\vspace{.05in}

\large{\bf Example 6:} Consider the resistive diffusion of an axial current filament\footnote{In a $Z$-pinch, the azimuthal magnetic field generated by an axial electric current in the plasma compresses and confines it.}, described by

\be\tag{56}
B_\theta\lp r,t\rp=\frac{I}{2\pi r}\lp 1\displaystyle-e^{-r^2/4\tilde{\eta}t}\rp
\ee

\noindent where $\tilde{\eta}\equiv\eta c^2$ and $I$ is the total axial current.

Using (56), (45a) gives

\be\tag{57}
\lambda^2=-\frac{I^2}{4\pi^2 r^3}\lp 1-e^{-r^2/4\tilde{\eta} t}\rp\lb-\frac{1}{r^2}\lp 1-e^{-r^2/4\tilde{\eta}t}\rp+\frac{1}{2\tilde{\eta} t}e^{-r^2/4\tilde{\eta} t}\rb
\ee

\noindent which leads to 

\be\tag{58}
\lambda^2=
\lc
\begin{matrix}
\begin{aligned}
\displaystyle-\frac{I^2}{64\pi\tilde{\eta}^2t^2},~&~r<<\sqrt{4\tilde{\eta}t}\\
\\
\displaystyle\frac{I^2}{4\pi r^4},~~&r>>\sqrt{4\tilde{\eta}t}
\end{aligned}
\end{matrix}\right.\ee

\noindent signifying elliptic region inside the pinch core and hyperbolic region away from the pinch core.

\vspace{0.3in}

\noindent\Large\textbf{6. Discussion}\\

\large Despite its extensive use as a diagnostic tool, the Okubo-Weiss criterion is primarily validated on empirical grounds by the results ensuing its applications, so some insight into the underlying connections is of much interest. In this direction, we have considered here recasting the Okubo-Weiss criterion by using the 2D hydrodynamic Beltrami condition (Shivamoggi et al. \cite{Shi}) that approximates the {\it slow flow-variation ansatz} imposed in its derivation.  This turns out to provide an interesting interpretation of the Okubo-Weiss criterion very logically in terms of the topological properties of the underlying vorticity manifold. Extension of these considerations to 2D quasi-geostrophic flows (via the {\it potential divorticity} framework) shows the robustness of the Okubo-Weiss parameter under the $\beta$-plane approximation to the Coriolis parameter. Inclusion of the $\gamma$-effect, however, produces changes in the Okubo-Weiss parameter. Extension to 2D MHD flows, on the other hand, provides one again with a useful diagnostic tool to parameterize the magnetic field topology in 2D MHD flows.
\vspace{.3in}

\noindent\Large\textbf{6. Acknowledgments}\\

\large BKS is thankful to Professor Jens Rasmussen for helpful discussions and would like to thank The Netherlands Organization for Scientific Research (NWO) for the financial support.

\end{document}